\documentclass[5p]{elsarticle}

\usepackage[switch]{lineno}
\modulolinenumbers[5]
\usepackage[hidelinks]{hyperref}

\newcommand{\dadd}[1]{#1}
\newcommand{\drem}[1]{}

\usepackage{amsmath}
\usepackage{bm}
\usepackage{units}
\usepackage{physics}
\usepackage{todonotes}
\usepackage{csquotes}

\journal{Physical Review D}

\begin{document}

\begin{frontmatter}

\title{Treatment of flux shape uncertainties in unfolded, flux-averaged neutrino cross-section measurements}

\author{Lukas Koch\corref{mycorrespondingauthor}}
\address{University of Oxford}

\cortext[mycorrespondingauthor]{Corresponding author}
\ead{lukas.koch@physics.ox.ac.uk}

\author{Stephen Dolan}
\address{CERN}

\begin{abstract}
The exact way of treating flux shape uncertainties in unfolded, flux-averaged neutrino cross-section measurements can lead to subtle issues when comparing the results to model predictions.
There is a difference between reporting a cross section in the (unknown) real flux, and reporting a cross section that was extrapolated from the \drem{(unknown)} real flux to a \dadd{well defined,} fixed reference flux. \drem{A lot of (most?)}\dadd{Many} current analyses do the former, while the results are compared to model predictions as if they were the latter.
This leads to \drem{(part of)} the flux shape uncertainty being ignored \dadd{at least partially},
potentially leading to wrong physics conclusions.
\dadd{A somewhat qualitative study of two results from T2K and MINERvA as examples suggests that the size of the effect is sub-dominant, but non-negligible for those measurements.}
\drem{The size of the effect is estimated to be sub-dominant, but non-negligible in two example measurements from T2K and MINERvA.}
This paper describes how the issue arises and \drem{provides instructions for possible ways how to treat the flux shape uncertainties correctly}\dadd{details possible methods for treating the flux shape uncertainties correctly}.
\end{abstract}


\end{frontmatter}

\section{Introduction}

Most modern neutrino cross-section measurements are reported as flux-averaged cross sections,
since the incident neutrino energy is not known on an event-by-event basis.
The average cross section $\sigma$ is reported as the number of expected events \drem{\footnote{\drem{within a certain kinematic and/or topological bin}}} $N$ per total incident neutrino flux $\Phi$ and number of targets $T$:
\begin{equation}
    \sigma = \frac{N}{T\Phi} \text{.}
\end{equation}
This reflects the capabilities of the detectors as closely as possible
and avoids making any assumptions about the neutrino energy dependence of the cross-sections one is trying to measure.
\dadd{This also applies for differential measurements, where $N$ only counts the number of events in certain kinematic and/or topological bins.}

The down-side of this approach is that each cross section is specific for its neutrino beam.
Even ignoring all smearing, efficiency, and acceptance issues,
the average cross section will change when the neutrino energy spectrum changes,
since the non-averaged cross sections depend non-trivially on the neutrino energy.
It is thus usually not possible to naively compare the results of different experiments with one another.
Instead one must use the theoretical energy dependence to calculate expected cross sections from interaction models for each flux separately.
This is all well understood and the process of generating the expected cross sections for multiple fluxes/experiments\footnote{including such annoyances as sensitive phase space constraints} has been made easier by frameworks such as NUISANCE~\cite{Stowell2017}.

The issue gets slightly more confusing when considering flux (and especially flux shape) uncertainties.
If one is not careful, it is easy to confuse two very similar ways of measuring and reporting the flux-averaged cross section:
\begin{enumerate}\label{eq:simple_xsec}
    \item Reporting the average cross section in the \emph{real} neutrino flux;
    \item Reporting the average cross section in a \emph{reference} neutrino flux.
\end{enumerate}
These two approaches lead to different uncertainty estimations and different rules when comparing the result to theoretical predictions,
\emph{even if} the reported central values for both methods are identical\footnote{e.g. when the chosen reference flux is the best fit value from a fit to the data}.

The first case is the conceptually easier one.
One takes the data and a detector model,
and then undoes the detector effects assuming a variety of detector, cross-section, and flux parameters.
For each set of parameters one calculates a flux-averaged cross section according to \autoref{eq:simple_xsec}.
The spread of results yields the reported uncertainty of the measurement.

The second case is slightly more complicated, as it contains an additional step.
After calculating\drem{\footnote{\drem{or as part of that calculation}}} the flux-averaged cross section in the real detector,
one needs to translate that result to the expected result in a \emph{fixed} reference flux.
I.e. the variation of the assumed real flux is used to vary the extrapolation from the data to the reference cross section.

The difference between the two approaches is subtle,
and both approaches yield correct results for what they are.
Problems arise when the result of one approach is treated like one of the other.

This becomes apparent when comparing the first kind of result with a theoretical prediction.
The predicted flux-averaged cross section is usually only calculated for a single flux,
and it is assumed that all flux uncertainties are contained within the covariance matrix of the published result, i.e. a second-approach result.
When the covariance matrix describes the uncertainties of the flux-average cross section in the \emph{real} flux, i.e. a first-approach result,
this is assumption is not correct.
This leads to \drem{part of} the flux shape uncertainty being ignored at least partially,
and possibly wrong conclusions to be drawn about the compatibility of a cross-section model with the data.
The following examples will illustrate this issue.

\subsection{Example A}
\label{sec:exampleA}

Imagine a very well though-out experiment that measures the total cross section of some neutrino interaction mode.
Let us assume that it is so well designed that efficiencies are perfectly flat \drem{(or even $100\%$)} for all relevant events, backgrounds are negligible, and systematic detector and model uncertainties play no role.
Also, the experiment has run for a very long time, so statistical errors are not an issue either.
This means the only uncertain term in \autoref{eq:simple_xsec} is the total neutrino flux $\Phi$.
An uncertainty in the total flux \drem{(let us assume $5\%$)} will be reflected in an equivalent relative uncertainty in the cross-section measurement.
\dadd{Let us assume a total flux uncertainty of $5\%$.}
Flux \emph{shape} uncertainties on the other hand will not be reflected in the result,
\dadd{as they do not affect the total flux $\Phi$.}
\dadd{For example, an uncertainty in the flux energy scale $\alpha$ modifies the assumed differential neutrino flux by proportionally shifting all energies up or down:
\begin{equation}
    \dv{\Phi}{E_\nu}{}(E_\nu, \alpha) = \dv{\Phi}{E_\nu}{}(E_\nu / \alpha, 1) / \alpha\text{.}
\end{equation}
It does not change the total integrated flux though, so
}
\drem{For example,} if there was e.g. a $15\%$ uncertainty on the flux energy scale,
\drem{this} \dadd{it} would not affect the cross-section result\drem{,
as the total number of neutrinos is not affected by this}.

Now imagine a model that predicts a cross section that is proportional to the neutrino energy.
The usual way of comparing the model to the measurement would be to use the nominal neutrino flux to calculate the total cross-section.
If the model predicts a $15\%$ lower cross section at the nominal flux, it could be ruled out at the 3~sigma level.
This does not consider the flux shape uncertainty though.
A variation of the neutrino flux energy scale by $15\%$ could easily explain the discrepancy, and would reduce the significance to below one standard deviation.

\subsection{Example B}
\label{sec:exampleB}

Let us now complicate things a bit by introducing a flux shape dependence in the experimental result.
The flux shape can enter \autoref{eq:simple_xsec} via the enumerator.
The expectation value of number of true signal events $N$ is rarely directly accessible, but a function of the number of recorded events $N_\text{rec}$, the estimated background contribution $N_\text{BG}$ and the assumed efficiency of signal event recording $\epsilon$\drem{\footnote{\drem{again ignoring statistical uncertainties}}}:
\begin{equation}
    N = \frac{N_\text{rec} - N_\text{BG}}{\epsilon} \text{.}
\end{equation}
\dadd{We are again assuming statistical uncertainties are negligible.}

Both the efficiency and the expected background can depend on the assumed flux shape.
The background is straight forward, as the effect of background processes depends on the flux of neutrinos at energies that contribute to them.
The efficiency can depend on the flux if the detector performance depends on event properties that change within the analysis bin,
and which in turn depend on the neutrino energy.
It is good practice to minimise efficiency uncertainties by choosing fine analysis bins, and by applying phase-space constraints to the signal definition.
The former is limited by the available statistics of the data and the detector resolution,
while the latter adds the excluded events to the background prediction,
so the flux dependence now enters via that route.\footnote{The flux dependence of the background prediction can of course be reduced by using data driven background subtraction methods, like the use of control regions.}

For this example, let us ignore the efficiency and concentrate on the background prediction.
Imagine a background process with a cross section \emph{anti}-proportional to the neutrino energy.
This means the \enquote{measured} number of true signal events will depend on the assumed neutrino flux energy scale.
The total uncertainty of the result will be larger than $5\%$,
depending on the background contribution in the recorded events.
Our test model still predicts $15\%$ too few events at the nominal flux.
With the included flux shape errors and thus increased total uncertainty, this is now less than a three-sigma effect.
When considering the flux shape uncertainty in the model prediction -- still at $15\%$ -- the significance of the difference again goes down below even one sigma.

This ignores the fact that the uncertainties of measurement and prediction are now correlated, though.
An assumed increase in real neutrino energy would increase the model prediction for a better fit to the data, but it would also \emph{decrease} the background contamination, increasing the \enquote{measured} number of signal events and thus the cross-section result.
Note that the direction and size of the correlation effect depends on the assumed background model.
In extreme cases, a correct treatment of the uncertainty could even \emph{increase} the significance of the model to data discrepancy compared to ignoring the flux shape uncertainty of the prediction.

\section{Challenges in the first approach}
\label{sec:1stapp}

The first approach -- reporting a cross section averaged over the \emph{real} flux -- is easier to measure in a model-independent way.
Depending on the complicating factors of a real-world experiment and the employed unfolding algorithms,
no assumptions about the neutrino energy dependence of the cross section need to be made (see e.g. \nameref{sec:exampleA}).

It does however make it challenging to compare the result with theoretical predictions.
Since the covariance matrices that go with such a result do not cover all effects of the flux on the model prediction,
it is necessary to calculate the model prediction uncertainty resulting from the uncovered flux errors.
This means that \dadd{the flux uncertainties must be propagated through any model in order for it to be correctly compared with the result}\drem{theorists will have to propagate the flux uncertainties through their models when doing a comparison}.

When doing so, another issue arises though.
The experimental results will have a flux contribution to their covariance matrix.
When the model predictions also gain a flux error,
those should be correlated to calculate correct goodness of fit scores.
Otherwise the correlations can lead to over- or under-coverage.

In some cases -- like \nameref{sec:exampleA} -- the flux uncertainty can be cleanly broken up into a part that affects the measured result, e.g. the normalisation, and a part that affects the model prediction, e.g. the shape.
In those cases it should be possible to have uncorrelated flux errors in data and model prediction.
In general this is not the case though,
and the flux uncertainty cannot be split into an experimental and a theoretical part (see e.g. \nameref{sec:exampleB}).
In those cases the correlations need to be taken into account
and the process of comparing a model to the data becomes a more involved statistical issue.%

Let us assume a cross-section measurement reports the point estimates for a vector of cross-section values $\bm{\hat{x}}$, a vector of flux parameters $\bm{\hat{\phi}}$, as well as the covariance matrix $S$.
In general, $S$ describes the uncertainties in the measurement and the correlations between cross-section bins and the flux parameters.
If the flux and cross-section uncertainties are uncorrelated,
this will simply mean that the respective off-diagonal elements of $S$ will be 0.\footnote{Conversely, if an experiment only provides two separate covariance matrices for the cross-section result and the flux parameters, one can only treat them as uncorrelated.}
Furthermore let us assume a cross-section model that can, given a set of flux parameter values, produce a prediction $\bm{m}(\bm{\phi})$ for the values of $\bm{x}$.

For an ideal likelihood ratio test of the model, we would like to calculate the log likelihood ratio
\begin{equation}
    -2\lambda = -2\sup_{\bm{\phi}}\qty[\ln(\Big.L(\bm{m}(\bm{\phi}), \bm{\phi}))]\text{,}
\end{equation}
where $\sup$ is the supremum function, maximising over $\bm\phi$, and
\begin{multline}
    -2\ln(L(\bm{x},\bm{\phi})) = \\
    \mqty*(\bm{x}^T - \bm{\hat{x}}^T & \bm{\phi}^T - \bm{\hat{\phi}}^T) S^{-1} \mqty*(\bm{x} - \bm{\hat{x}} \\ \bm{\phi} - \bm{\hat{\phi}})
\end{multline}
describes the likelihood surface of the parameter space. If the model is true, $-2\lambda$ will be $\chi^2_k$-distributed with $k$ equal to the number of bins in $\bm{x}$.
The usual quantiles of $\chi^2_k$ can then be used to judge the goodness of fit of the model to the data.

Unfortunately, depending on the complexity of calculating $\bm{m}(\bm{\phi})$, it might not be feasible to maximise $\lambda$ over the flux parameter space.
In that case one can use the following likelihood ratio as approximation:
\begin{equation}
\begin{split}
    -2\lambda' &= -2\ln(\Big.L'(\bm{m}(\bm{\hat\phi}))) \\
    &=\qty(\bm{m}^T(\bm{\hat{\phi}}) - \bm{\hat{x}}^T) S'^{-1} \qty(\bm{m}(\bm{\hat{\phi}}) - \bm{\hat{x}})\text{.}
\end{split}
\end{equation}
Here $S'$ is \emph{not} just the cross-section part of $S$.
It is a new covariance matrix that describes the expected variation of the distribution of $(\bm{m}^T(\bm{\hat{\phi}}) - \bm{\hat{x}}^T)$ and thus $-2\lambda'$, assuming that the model is correct.
It can be calculated from $S$, $\bm{\hat{x}}$, $\bm{\hat\phi}$, and $\bm{m}(\bm\phi)$ by generating random samples of $\bm{x}$ and $\bm{\phi}$ and then calculating
\begin{equation}
    \bm\Delta = \qty(\bm{m}(\bm{\phi}) - \bm{x})
\end{equation}
for each sample.
The sample covariance of $\bm\Delta$ can then be used as an estimator for $S'$.
If there is no correlation between the flux and cross-section parameters, this procedure is equivalent to calculating the covariance of $\bm{m}(\bm\phi)$ only and adding it to the cross-section part of $S$.

Note that we are ignoring the sample mean of $\bm\Delta$.
We calculate a distribution of differences between data and prediction at the best-fit point of the data and then apply the resulting variance to the model prediction.
This approach makes two assumptions:
\begin{itemize}
    \item The covariance matrix $S$, describing the likelihood surface of the cross-section and flux parameters, can be interpreted as the covariance of the expected spread of maximum likelihood estimators (MLEs).
    \item The covariance of the MLEs is constant in the relevant parameter range, while the expectation value depends on the true parameter value.
\end{itemize}
These assumptions should hold well enough for all measurements that make a Gaussian error approximation -- as is implied by reporting the uncertainty as a covariance matrix.
The second assumption might break if the model prediction $\bm{m}(\bm\phi)$ and the best fit data result $\bm{\hat{x}}$ are very different.
But this should not change the physics conclusions of a model comparison.\footnote{At some point it is no longer important whether the data to model agreement is bad or terrible.}
If all assumptions hold and the model is actually true, the test statistic $-2\lambda'$ should again be $\chi^2_k$ distributed and the usual critical values apply.

\section{Challenges in the second approach}
\label{sec:2ndapp}

The second approach -- reporting a cross section averaged over a \emph{reference} flux -- makes it easier to compare the results to model predictions.
The covariance matrix already includes by construction the uncertainties of extrapolating the data from the unknown real neutrino flux to the reference flux.
When testing a model, one only needs to calculate the flux-averaged cross section in that reference flux.

It is however more difficult to produce a result in this manner,
because the extrapolation to a reference flux is not trivial.
It necessarily requires a cross-section model to do so,
since only a model can predict the connection between true neutrino energy and measured variables\footnote{Except when measuring a cross section in terms of neutrino energy directly, of course. But in that case this whole paper becomes a moot point.}, e.g. neutrino kinematics, particle multiplicities, etc.

So the extrapolation itself -- and with it the cross-section result -- will be subject to cross-section model uncertainties.
This somewhat undermines the aim of producing a model-independent measurement.
If the tested model was not covered by the assumed model-uncertainties -- i.e. if it requires a different propagation of flux shape uncertainties compared to the considered models --
the method of comparing the model with the data at only the reference flux breaks down again.

To illustrate this, consider \nameref{sec:exampleA}.
If the model predicting a cross section proportional to the neutrino energy was used to extrapolate to the reference flux,
the total uncertainty on the result would be about $16\%$.
Now consider a second model, which predicts a cross section proportional to the square of the neutrino energy
and over-predicts the measured value by $50\%$.
With the reported uncertainty, this would be considered a three-sigma effect,
but the squared neutrino energy dependence means
that the $15\%$ energy scale uncertainty should translate to a $30\%$ rate uncertainty.
Again the model is more compatible with the data then the covariance matrix implies.

This suggests that one should be conservative and cover as many possible flux propagation models as possible.
Unfortunately this also degrades the power of the measurement.
Imagine a third model that predicts a cross section that is constant over all neutrino energies, and overestimates the data of \nameref{sec:exampleA} by $15\%$.
With the assumed flux shape uncertainties, this would not even be a one sigma difference.
But when considering that the model does not predict any shape dependence,
it should be refuted at the three sigma level from the $5\%$ flux normalisation uncertainty.

Overall, the feasibility of the second approach will depend on the size of the flux shape uncertainty and how well constrained the neutrino-energy dependence of the measured cross section is.
If the effect is large, and the energy dependence is uncertain, it might be better to opt for a first-approach measurement instead.

\section{Size of the effect}
\label{sec:sizeOfEffect}
Whichever approach is taken, the effect we consider here comes from the flux shape dependence of the cross-section prediction.
In order to assess the relevance of this effect for modern cross-section measurements, the spread of flux-averaged cross section predictions are evaluated for a fixed model propagated through an ensemble of flux predictions.
These flux predictions are constructed from T2K~\cite{Abe:2012av} and MINERvA~\cite{Aliaga:2016oaz} nominal fluxes and accompanying covariance matrices\footnote{The MINERvA flux extends up to $\unit[100]{GeV}$ but, the numerical precision of the provided covariance matrix makes the complete matrix non-invertible. Here we consider only the first $\unit[35]{GeV}$ (containing $\sim99.4\%$ of the flux). It should also be noted that the T2K covariance is provided in much coarser bins than the accompanying flux prediction.}.
Whilst the flux prediction is varied in both shape and normalisation, the latter plays no role in constructing a flux-averaged cross section from a model prediction.
The model used is provided by GENIE~\cite{Andreopoulos:2009rq} (Version 3.00.06, tune G18\_10b\_00\_000).
The specific cross sections that are predicted are MINERvA's CCQE-like measurement from Ref.~\cite{Ruterbories:2018gub} and T2K's CC0$\pi$ Analysis~I in Ref.~\cite{Abe:2016tmq}.
Both of these cross sections are double-differential in outgoing muon kinematics.
MINERvA measures the momentum broken down into its transverse ($p_{T}$) and longitudinal ($p_{||}$) components, defined with respect to the incoming neutrino direction, whilst T2K measures the magnitude of the momentum ($p_{\mu}$) alongside the cosine of the angle between the muon and neutrino ($cos(\theta_\mu)$).
Both measurements also follow the first approach, i.e. no attempt was made to extrapolate the measured cross section to a particular reference flux shape.
The model is compared to the cross sections using the NUISANCE framework~\cite{Stowell2017}.

\begin{figure*}
\centering
\includegraphics[width=0.49\textwidth]{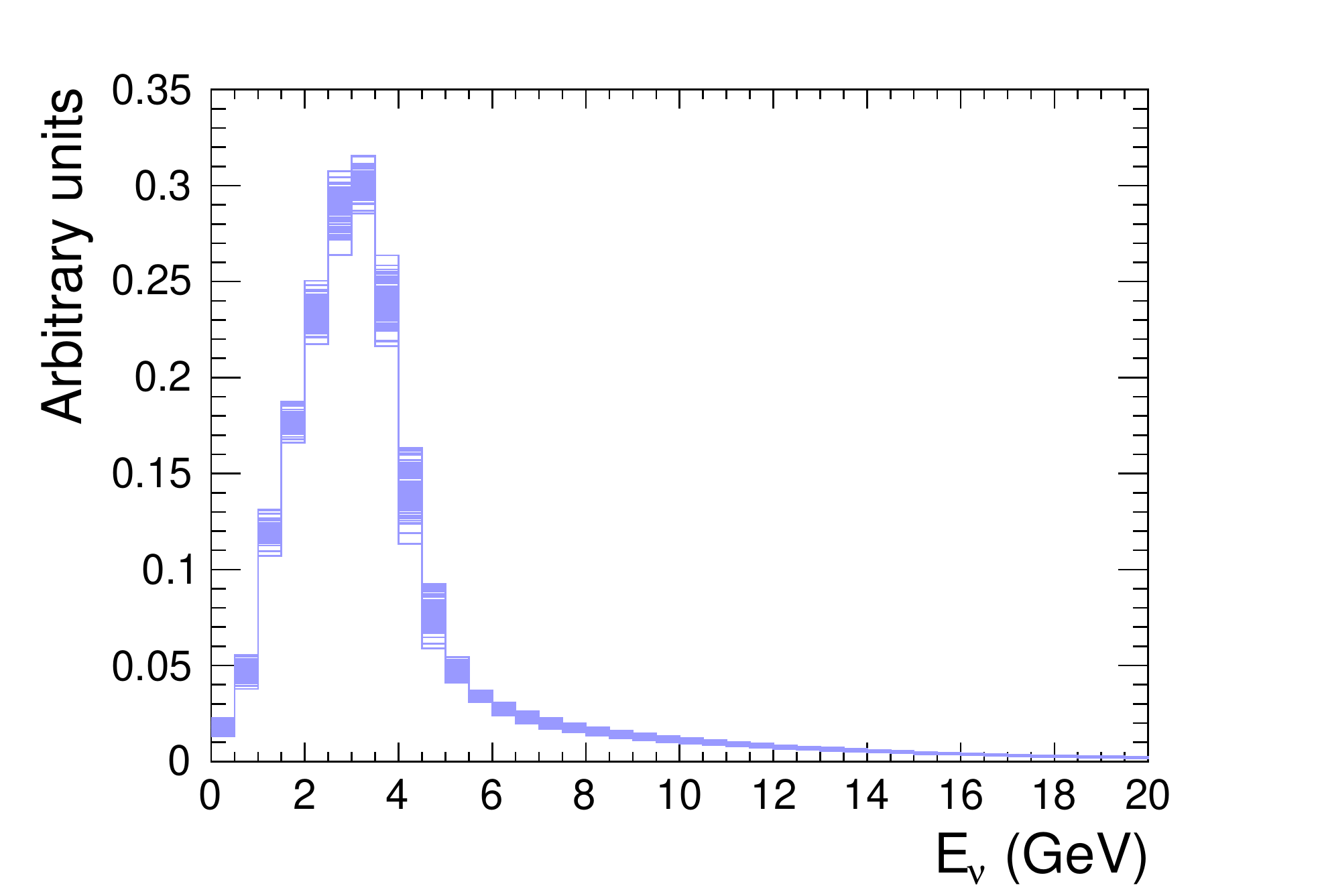}
\includegraphics[width=0.49\textwidth]{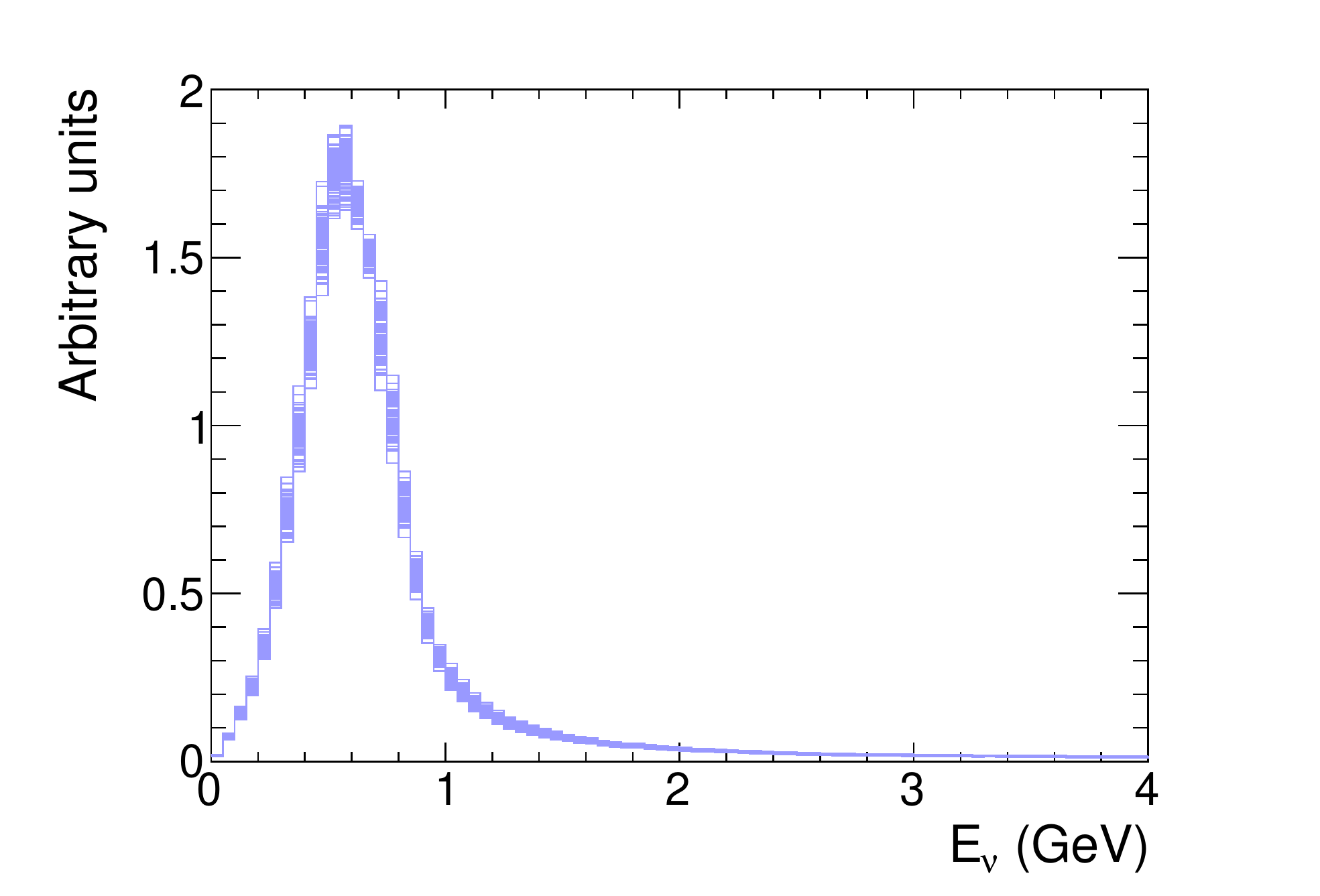}
\caption{\label{fig:fluxToys} Histograms showing the ensembles of flux predictions produced for MINERvA (left) and T2K (right) which have each been re-normalised to have the same integral. Note that the actual flux predictions extend up to 35 GeV for MINERvA and 10 GeV for T2K but for readability the long tails are not shown.}
\end{figure*}

\begin{figure*}
\centering
\includegraphics[width=0.49\textwidth]{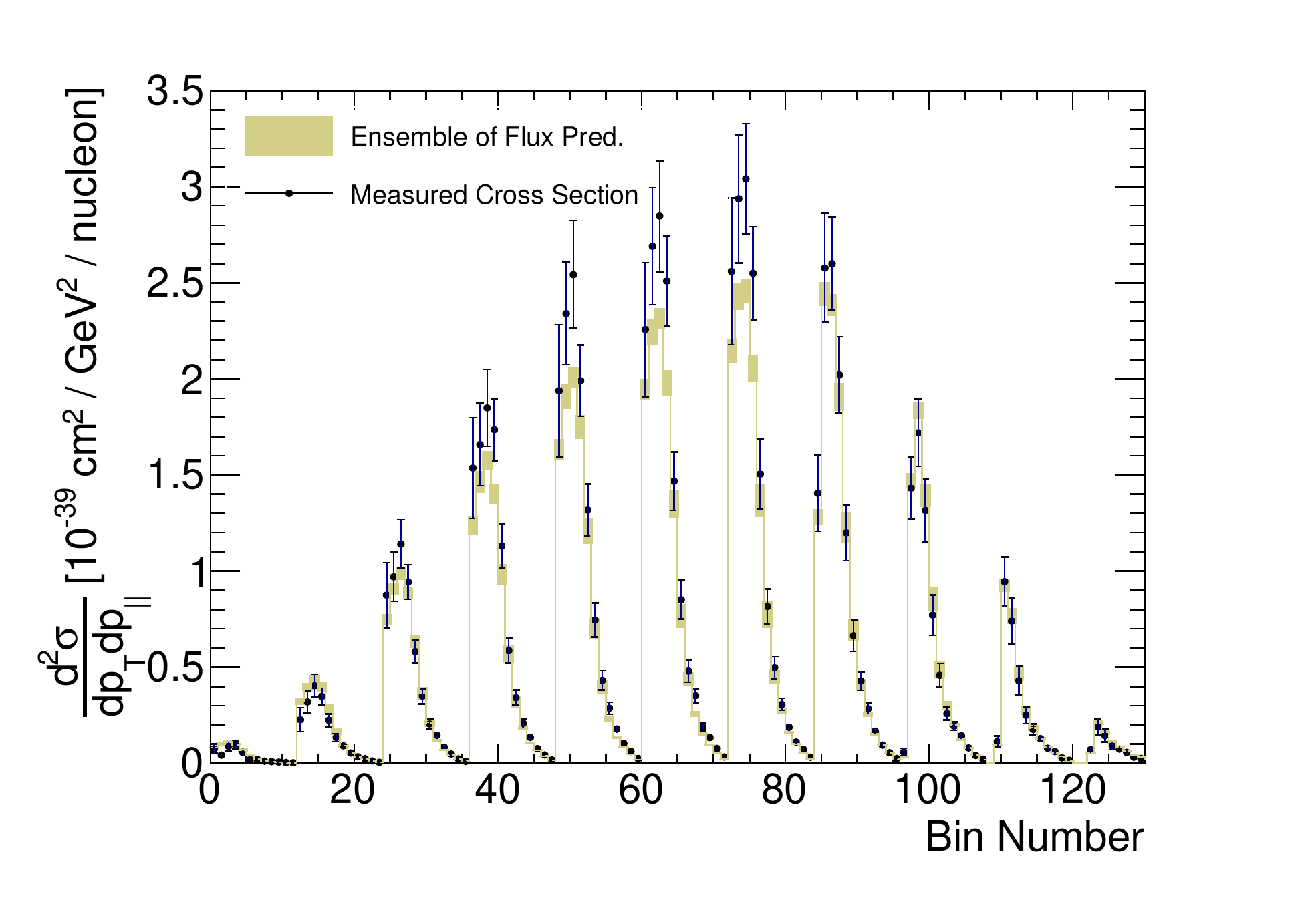}
\includegraphics[width=0.49\textwidth]{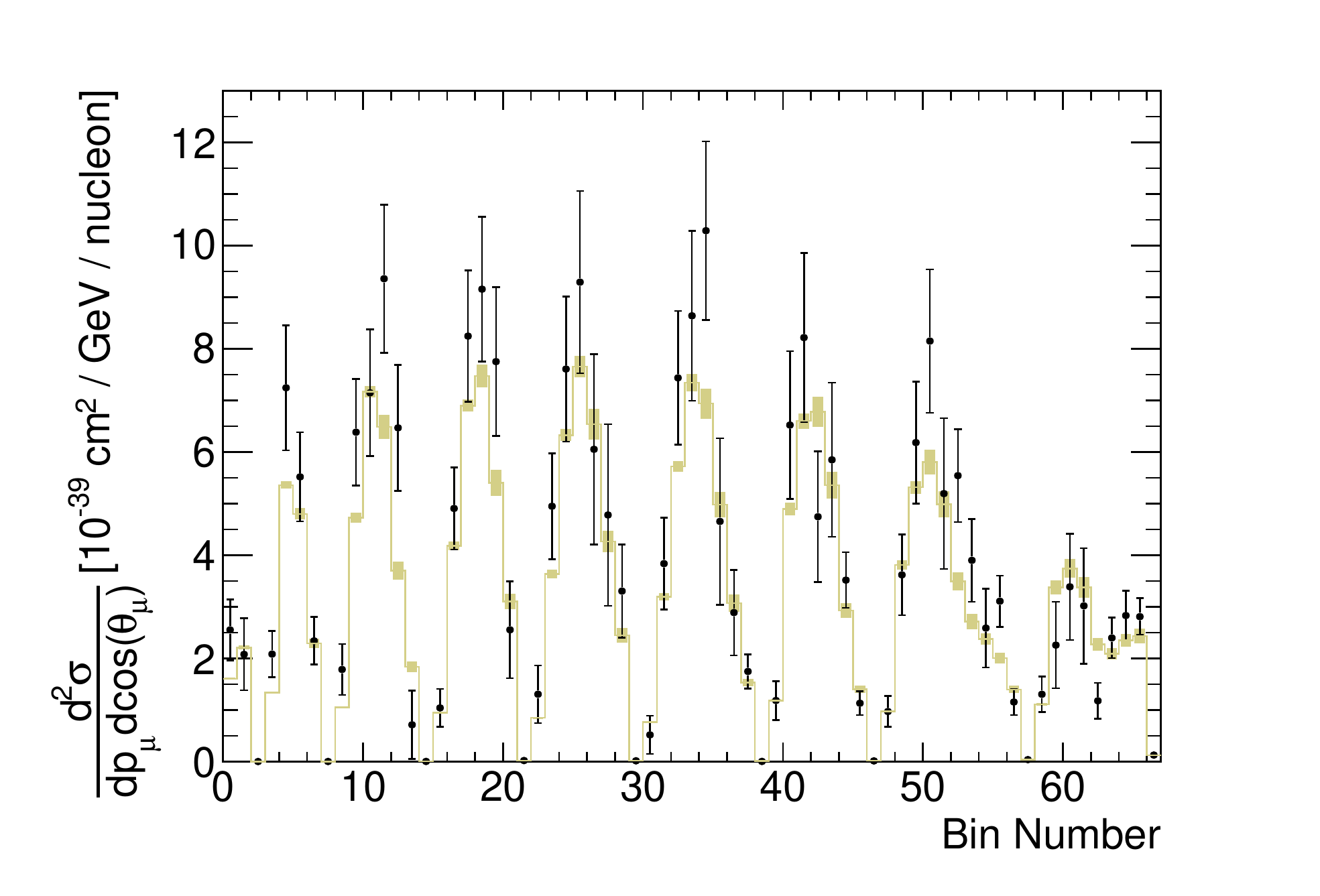}
\includegraphics[width=0.49\textwidth]{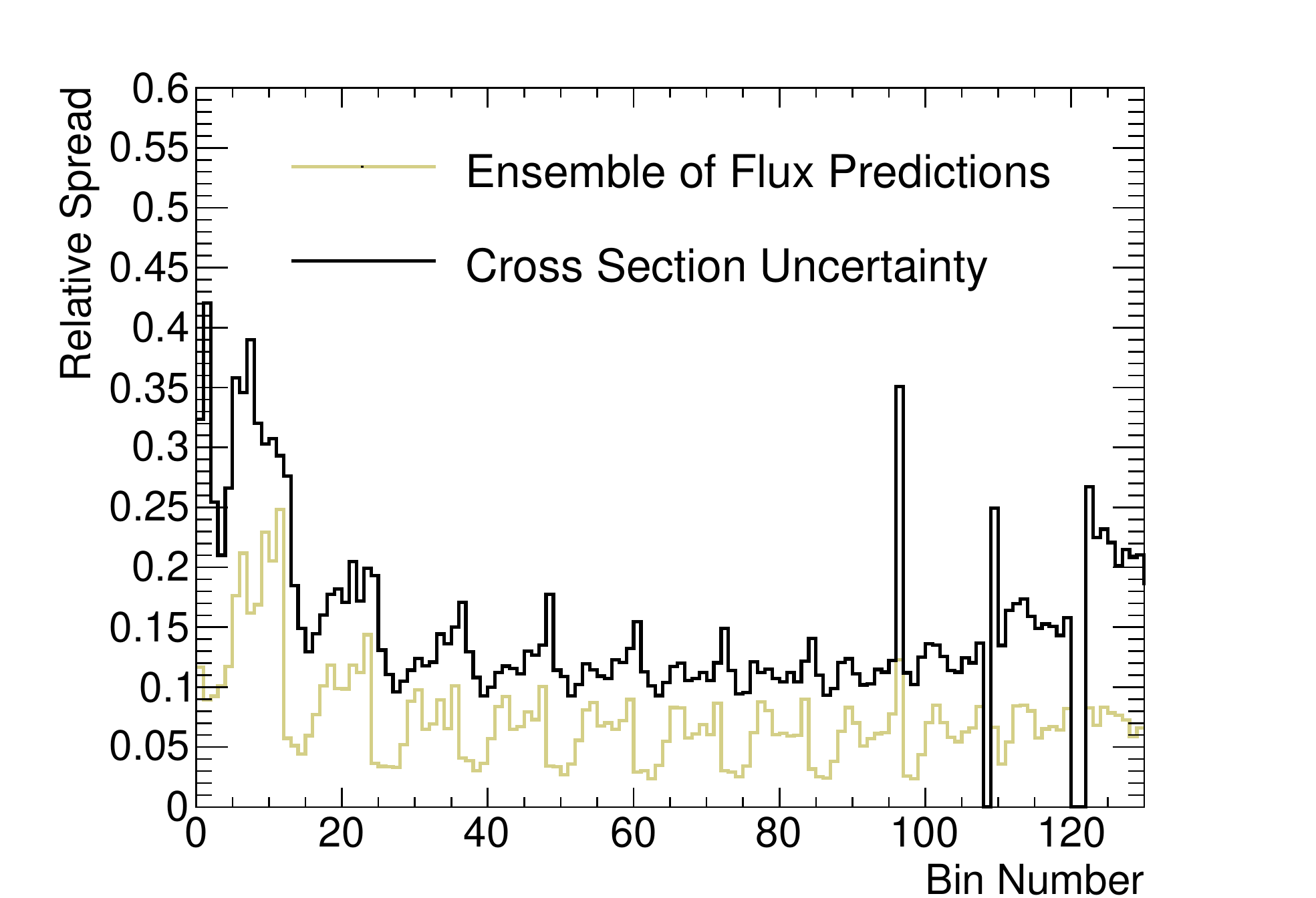}
\includegraphics[width=0.49\textwidth]{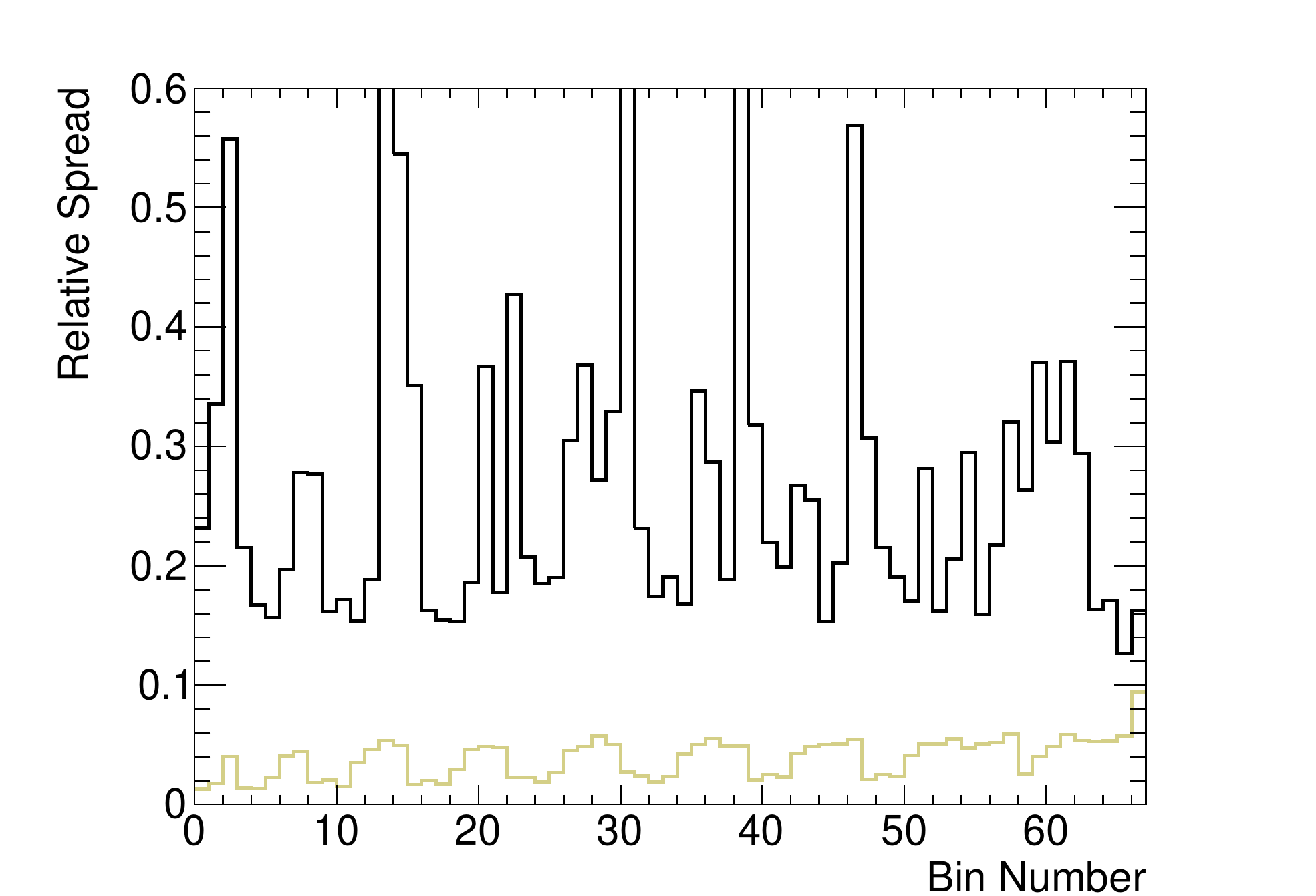}
\caption{\label{fig:difHisto}
\textbf{Top:} The differential cross sections measured by MINERvA (left) and T2K (right) are shown alongside the GENIE model prediction with an error band taken from the spread (standard deviation) of the predictions generated with the ensemble of flux predictions.
\textbf{Bottom:} The relative uncertainty (i.e. 0.2 means 20\% uncertainty) on the cross section measured by MINERvA (left) and T2K (right) is compared to the spread in the model predictions.
The x-axis in each plot is simply a bin number.
For MINERvA the bins are ordered in increasing $p_{||}$ in slices of increasing $p_{T}$.
For T2K the bins are ordered in increasing $p_{\mu}$ in slices of increasing $cos(\theta_\mu)$.
The very forward high momentum bins discussed in the text correspond to around bin 10 for MINERvA and the final few bins for T2K.
The MINERvA analysis has additional high-$p_{T}$ bins which are not shown for readability since they contribute a negligible proportion of the cross section.
}
\end{figure*}

The sampled ensembles of flux predictions are shown in \autoref{fig:fluxToys}.
The standard deviation of the mean \drem{flux} \dadd{neutrino energies} across the ensemble \dadd{$\text{std}(\ev{E_\nu})$}
is $\sim\unit[5.5]{MeV}$ for T2K and $\sim\unit[6.5]{MeV}$ for MINERvA.
However, it should be noted that this energy scale uncertainty is not fully representative of the flux shape uncertainties.
The spread of the cross-section predictions is shown in \autoref{fig:difHisto} alongside the cross-section measurements from the experiments.
It also shows a bin-by-bin comparison of the spread of the cross-section predictions compared to the uncertainty on the measurements.
\drem{Assuming no} \dadd{Ignoring all} correlations, this gives a \dadd{qualitative} indication how much of the error budget of the measurement is consumed by the flux shape prediction uncertainty,
\dadd{despite not being correctly included in the error propagation of the results.
I.e., this is a rough measure for the \enquote{missing} uncertainty when treating these first-approach measurements like second-approach ones.}
\drem{The spread of the cross-section predictions is exactly the additional uncertainty that must be included in a model comparison for a first approach measurement if correlations between the ensemble of flux predictions and the flux uncertainty that is included in the cross-section measurements are neglected.
It also broadly corresponds to the the size of the missing uncertainty in a second approach measurement, although a complete second approach is more complicated and requires the addition of the flux-shape uncertainty at the time of the cross-section extraction (see Sec.~\ref{sec:recipes}).}

Although the spread in the prediction of the total cross section integrated across all bins is about 1\% for both MINERvA and T2K, the spread in individual bins of the differential cross section is much larger.
This is especially the case for bins corresponding to forward going muons with large momenta which are produced only by interactions of neutrino in the tail end of the flux predictions, where the flux uncertainty is largest.
In these bins, the spread of the \drem{flux} \dadd{cross-section} predictions are sometimes comparable to the entire uncertainty in the measured cross section.
This said, it should of course be noted that the uncertainty on each cross-section result is characterised by a full covariance matrix describing the correlations between bins rather than just the error bars.
It is therefore difficult to quantitatively compare the size of the model prediction variations with the variations allowed by the cross section uncertainty from only \autoref{fig:difHisto}.

\begin{figure*}
\centering
\includegraphics[width=0.49\textwidth]{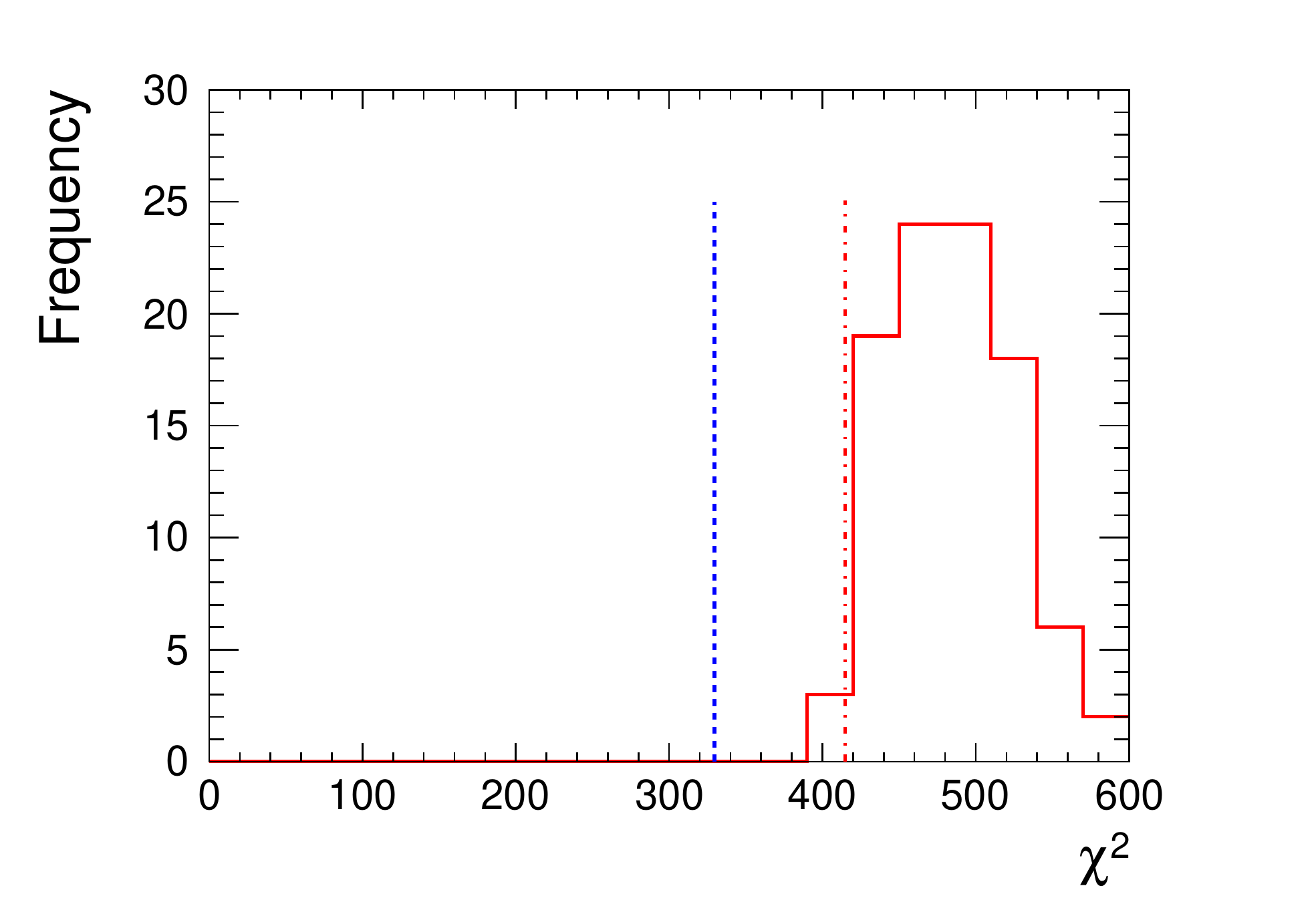}
\includegraphics[width=0.49\textwidth]{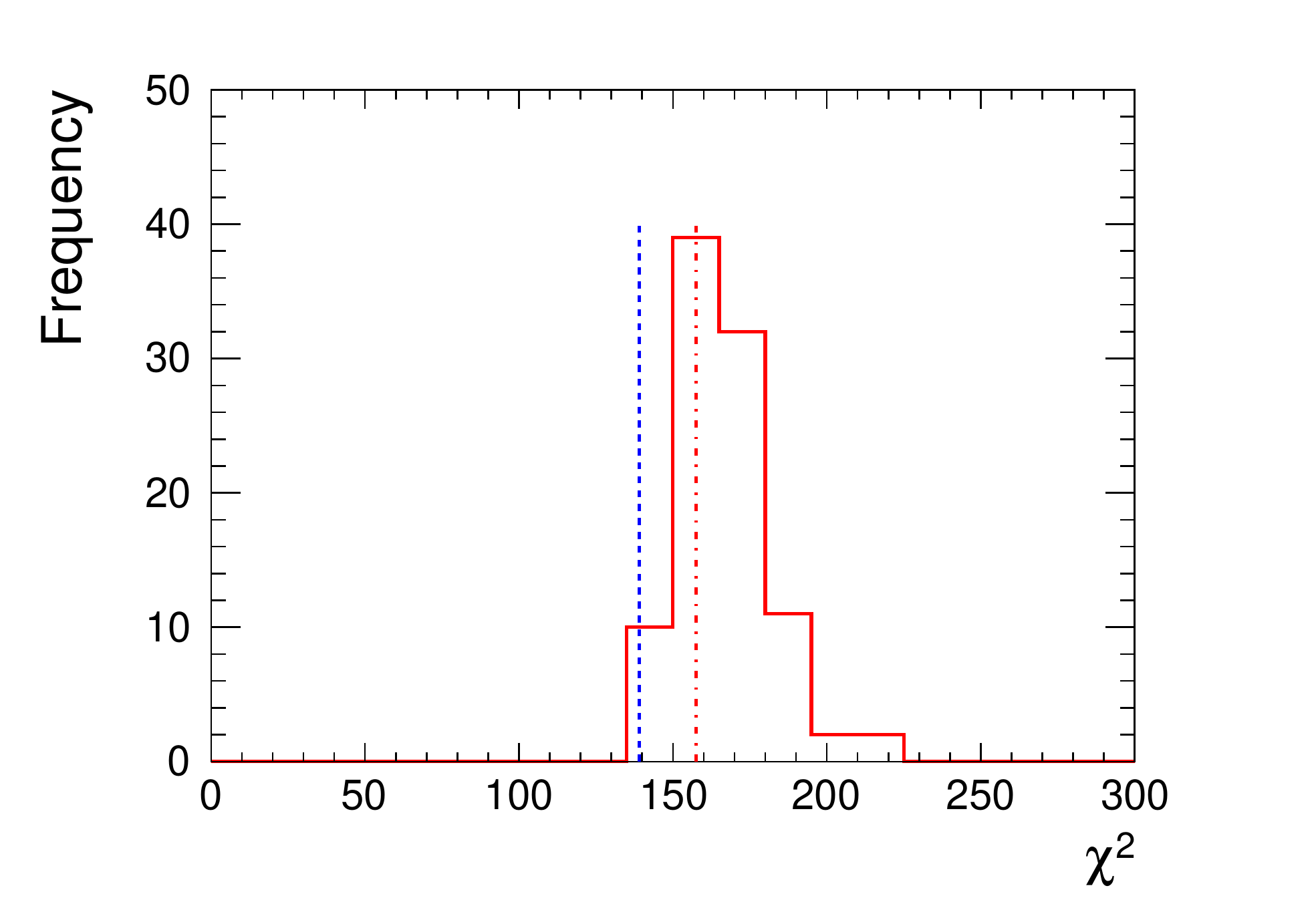}
\caption{$\chi^2$ comparison.\label{fig:chi2Comp}
The \enquote{original} $\chi^2$ calculated using the GENIE prediction using the nominal flux and the original covariance matrix is shown by the dot-dashed red line for MINERvA (left, 144 non-zero cross-section bins) and T2K (right, 67 cross-section bins).
The distribution indicated by the solid red line shows the same but with a different model prediction for each varied flux prediction.
The dashed blue line shows the $\chi^2$ from the comparison of the nominal prediction with the data using the modified covariance matrix, corresponding to the $-2\lambda'$ statistic introduced in \autoref{sec:1stapp}.
}
\end{figure*}

A more quantitative analysis of the relevance of the flux shape uncertainty can be performed by assessing how the $\chi^2$ statistic, calculated using the full covariance matrix to describe the uncertainty on the cross-section measurements, changes across the ensemble of flux predictions. \dadd{This $\chi^2$ is defined as:
\begin{equation}
    \chi^2 = (\bm{x}_{\text{MC},i} - \bm{x}_\text{data})^T S^{-1}_{\text{data}} (\bm{x}_{\text{MC},i} - \bm{x}_\text{data}) \text{,}
    \label{eq:chi2}
\end{equation}
where $S_{\text{data}}$ is the covariance matrix as reported by the experiments\footnote{Note that 12 of the 156 cross-section bins in the MINERvA data release contain zero measured cross section with zero uncertainty to reflect the kinematic acceptance of the detector. The resultant covariance matrix is therefore non-invertible. To calculate a $\chi^2$ statistic we followed the same approach as MINERvA and formed a pseudo-inverse using a Singular Value Decomposition (SVD) approach.}, $\bm{x}_{\text{MC},i}$ is the predicted cross section in the $i$\textsuperscript{th} flux shape variation and $\bm{x}_\text{data}$ is the cross section measured by the experiments.} \autoref{fig:chi2Comp} shows that this $\chi^2$ is much larger than the number of cross-section bins and so it is therefore hard to interpret beyond stating that all of the predictions from the ensemble of fluxes are in very poor agreement with the measurements.
This would be the case for the majority of currently available cross-section models.

As discussed in \autoref{sec:1stapp}, if potential correlations between the cross-section uncertainty and the spread of predictions from the ensemble of fluxes are neglected, an approximate combined covariance matrix can be formed by adding the covariance of the ensemble of predictions to the one cross-section measurement covariance.
This $\chi^2 = -2\lambda'$ -- calculated using the combined covariance -- is also shown in \autoref{fig:chi2Comp}.
Whilst it is clear there is a notable shift to lower $\chi^2$ when the combined uncertainty is considered, the model remains absolutely disfavoured by the measurements.

\begin{figure*}
\centering
\includegraphics[width=0.49\textwidth]{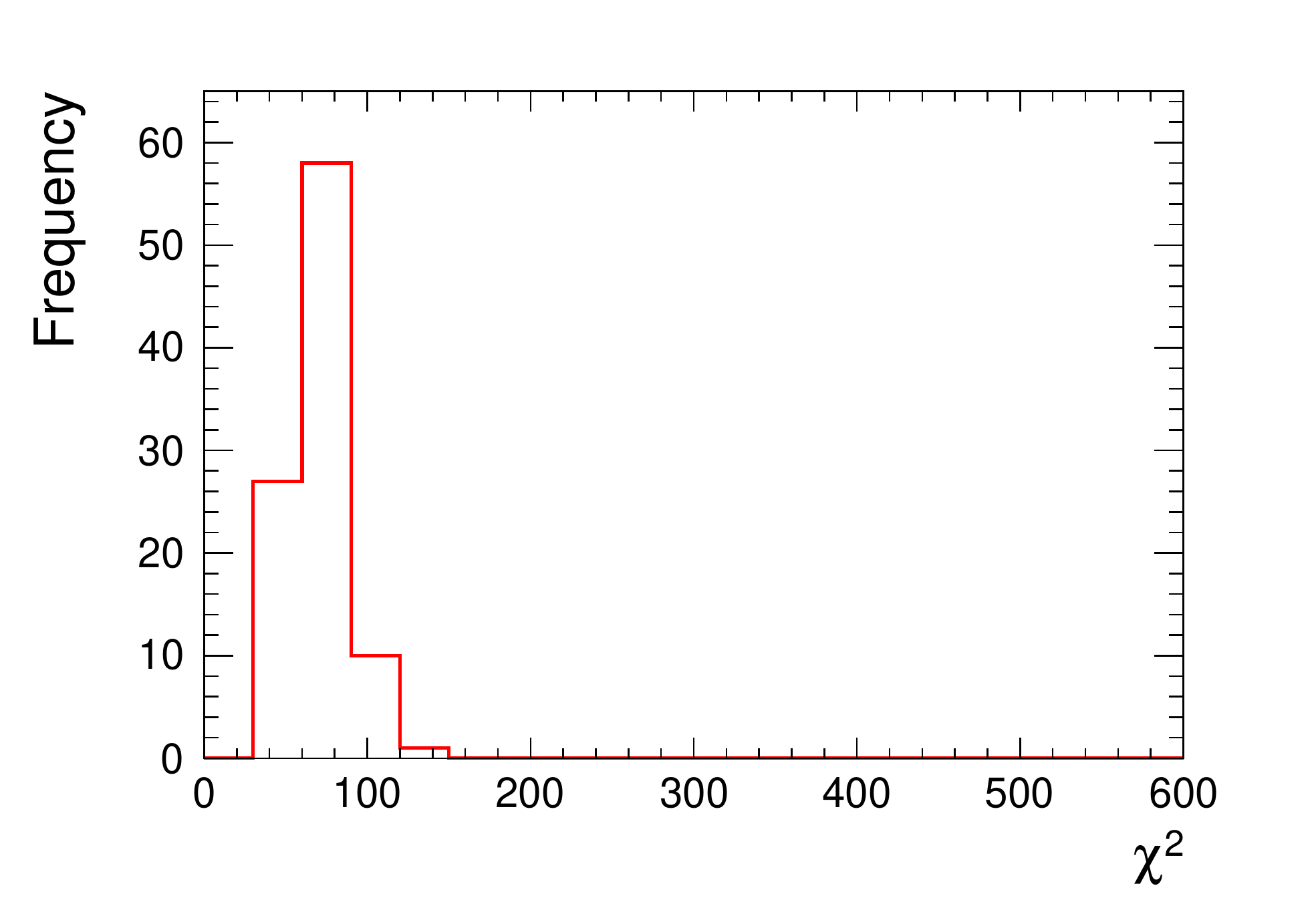}
\includegraphics[width=0.49\textwidth]{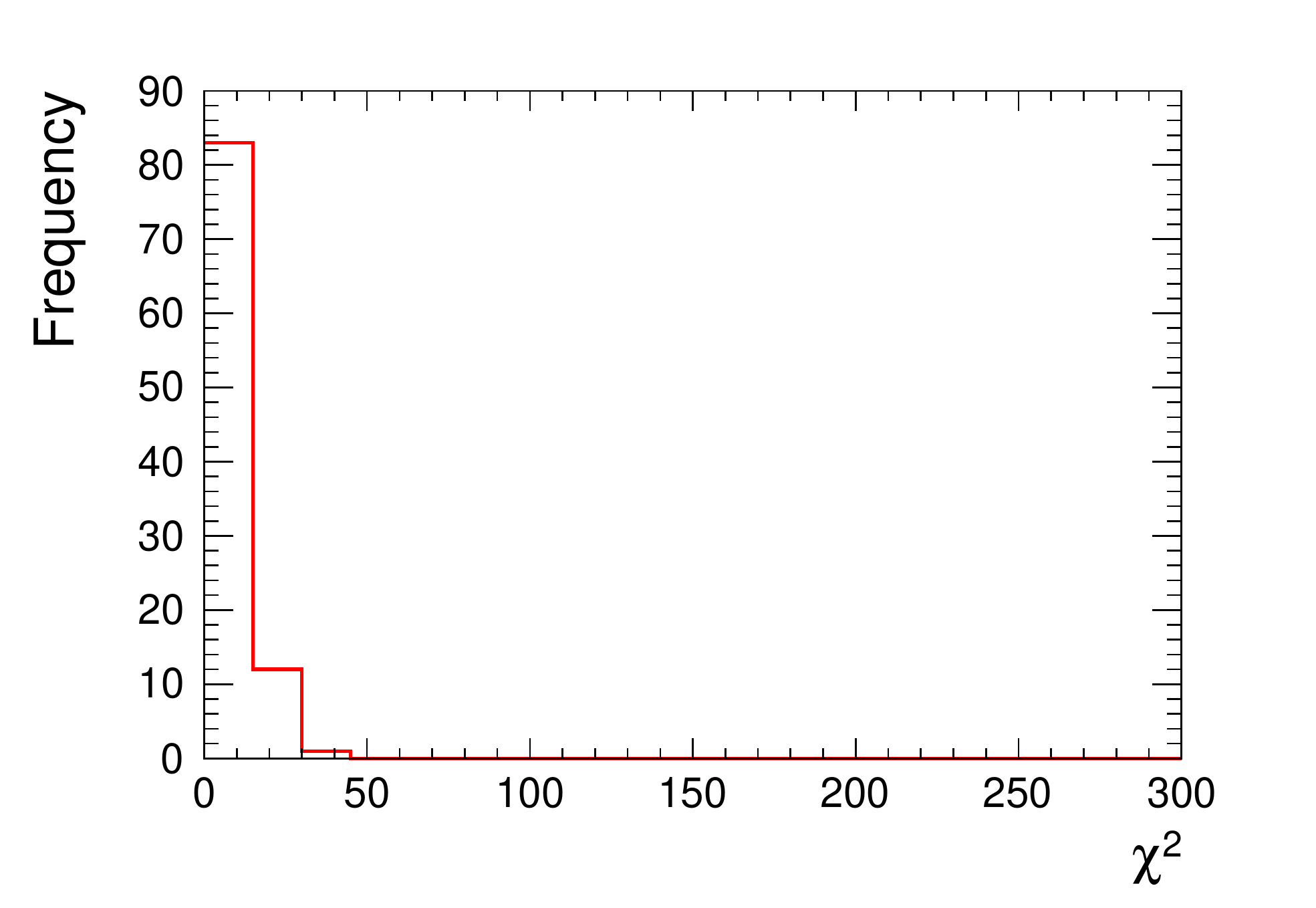}
\caption{\label{fig:chi2Mean}
The $\chi^2$ calculated for each of \dadd{flux-shape-}varied model predictions compared to the nominal \dadd{flux} GENIE prediction of the MINERvA (left, 144 non-zero cross-section bins) and T2K (right, 67 cross-section bins) cross-section measurements. The $\chi^2$ are calculated using the original covariance matrices provided by T2K and MINERvA.
}
\end{figure*}

An easier to interpret assessment of the relevance of the flux shape uncertainty compared to the rest of the cross-section uncertainty can be constructed by considering a scenario in which the T2K and MINERvA analyses had measured exactly the GENIE prediction with the nominal flux.
\dadd{For simplicity this scenario assumes that the covariance of the data remains unchanged which, whilst unrealistic, is a suitable approximation for this illustrative example.}
\drem{In this case, the spread of the $\chi^2$ calculated across the ensemble of flux predictions using the cross-section uncertainty covariance matrix directly}
\dadd{In this scenario, the spread of the $\chi^2$, calculated as in \autoref{eq:chi2} but where $\bm{x}_\text{data}$ is replaced by the GENIE model prediction for the nominal flux, should indicate more directly} the relative importance of the flux shape uncertainty.
This is shown in \autoref{fig:chi2Mean}.
The fact that this $\chi^2$ is almost always less than the number of cross-section bins shows that the uncertainty from the flux shape is in a sense \enquote{covered} by the other uncertainties on the cross section.
However, the change in the $\chi^2$ is clearly large enough such that conclusions regarding the suitability of slightly altered model predictions could be significantly changed if the flux shape variations were not to be considered.
Note that this study also continues to neglect potential correlations between the ensemble the flux predictions and the cross-section covariance.
\dadd{This should not affect the qualitative conclusions by much though,
since the flux shape contribution to the uncertainty of the cross-section results is relatively small.}
Overall the relative size of the flux shape uncertainty appears \dadd{non-negligible} \drem{important} but subdominant in the analyses considered.
\dadd{A more quantitative statement would require more in-depth studies in the experiments, following the recipes from \autoref{sec:recipes}.}

\section{Recipes}
\label{sec:recipes}

The following sections will describe how to implement the two approaches within certain unfolding and error propagation schemes.
This is not intended as an endorsement of these schemes,
nor can these instructions be blindly applied to different algorithms.
They should, however, provide examples of how to approach the issue in general.

\subsection{Template fitter}

This section concerns the template fitting approach of unfolding as e.g. described for the \enquote{STV} analysis in~\cite{Abe:2018pwo}.
In short:
\begin{itemize}
    \item Interaction models and detector simulations are used to create a prediction function of reconstructed events $N_\text{reco}$ as a function of multiple parameters, e.g. by re-weighting.
    \item A fitting algorithm is used to determine which combination of parameter values is or is not compatible with the real data.
    This information could be encoded in a best-fit point plus a covariance matrix of the parameters, for example.
    \item Sets of compatible parameter values are used to calculate the desired cross sections.
    The spread of compatible value sets translates to a spread of cross-section results,
    which e.g. can be parameterised as another covariance matrix.
\end{itemize}

The choice of parameters to vary and how to implement that variation is a complicated topic and outside the scope of this paper.
Let us just split the parameters into parameters that affect the assumed neutrino flux, the flux parameters $\bm\phi$, and parameters that do not $\bm\theta$.
The latter will usually contain parameters for the detector response, the (background) cross-section models, and the primary aim of the measurement: the template weight parameters for the measured cross section,
\begin{equation}\label{eq:xsec-parameters}
    \sigma = \frac{N(\bm\theta,\bm\phi)}{T(\bm\theta)\Phi(\bm\phi)} \text{.}
\end{equation}
Here $N$ is the predicted number of \emph{true} events.\footnote{For simplicity's sake, we will only consider total cross sections here. All arguments apply equally to differential measurements though. In fact, the calculations should be identical modulo a division by the areas of the analysis bins.}
Depending on the fitting algorithm this is either a direct output of the fit,
or it can be calculated from the number of predicted \emph{reconstructed} events via a predicted\footnote{Ensuring that the efficiency does not depend too much on any assumed cross-section or flux model is another can of worms that shall remain unopened in this paper.} efficiency $\epsilon$:
\begin{equation}\label{eq:efficiency}
    N(\bm\theta,\bm\phi) = \frac{N_\text{reco}(\bm\theta,\bm\phi)}{\epsilon(\bm\theta,\bm\phi)}\text{.}
\end{equation}

Like this, the method produces a first-approach result.
The propagated sets of parameter values correspond to possible \emph{real} experimental conditions,
and thus the calculated cross sections do the same.
It is easy to imagine that under ideal experimental conditions,
the variation of $N(\bm\theta,\bm\phi)$ will be small as the fitter ensures that the allowed parameter values fit the number of recorded events, which is a constant number.
In this case, the flux uncertainty will affect the cross-section result solely via the total flux $\Phi(\bm\phi)$.
Any shape uncertainty would have only a small effect via $N$, or no effect at all.

At this point it is also fairly easy to calculate a covariance matrix that correlates the cross-section result with the flux parameters.
Instead of only calculating the cross-section values for each throw of parameters, one also treats the thrown flux parameters themselves as part of the result and calculates a covariance matrix over the combined cross-section values and flux parameter values.

To extract a second-approach result from the fitter,
one needs to extrapolate the results of a set of parameter values to a reference flux.
We can specify the reference flux using a fixed set of flux parameters: $\bm\phi'$.
Let us call the extrapolation of expected events at that flux $N'$.
The calculated cross section then becomes
\begin{equation}
    \sigma' = \frac{N'(\bm\theta,\bm\phi)}{T(\bm\theta)\Phi(\bm\phi')}\text{.}
\end{equation}
It might seem counter-intuitive at first to fix the flux normalisation to the $\bm\phi'$ parameter values in this formula,
when the aim is to include \emph{more} of the flux uncertainty in the result.
This is not a contradiction though, since the flux uncertainty now enters the result via the extrapolation in $N'$.

The exact procedure will again depend on the flux parameters,
but let us assume the parameters are -- or can be converted to -- event weights binned in neutrino energy.
In this case, $N'$ can be expressed as
\begin{equation}\label{eq:extrapolation}
    N'(\bm\theta, \bm\phi) = \sum_i N_i(\bm\theta, \bm\phi) \frac{\phi'_i}{\phi_i} \text{,}
\end{equation}
with the neutrino energy bin index $i$, and the number of true events in each bin $N_i$.
In this form it is clear how the flux uncertainty enters the result.
Under ideal conditions, the fitter output will ensure that $\sum_i N_i(\bm\theta,\bm\phi)$ is compatible with the number of observed events by correlating the flux weights and the other parameters,
i.e. $N_i$ will not vary much under allowed parameter variations.
Then the ratio of the varied weights $\bm\phi$ and the constant reference flux weights $\bm\phi'$ modifies that number according to the flux uncertainty.

We can further separate the flux weights from the true event prediction:
\begin{equation}
    N_i(\bm\theta,\bm\phi) = N_i(\bm\theta)\phi_i \text{,}
\end{equation}
where $N_i(\bm\theta)$ is the number of predicted true events in the $i$-th energy bin in an unweighted, nominal\footnote{not necessarily the same as the reference!} flux.
So \autoref{eq:extrapolation} becomes
\begin{equation}\label{eq:full-extrapolation}
    N'(\bm\theta, \bm\phi) = \sum_i N_i(\bm\theta) \phi_i \frac{\phi'_i}{\phi_i} = N(\bm\theta,\bm\phi') \text{.}
\end{equation}
This means, the extrapolation to the reference flux happens by simply ignoring the thrown flux weights and calculating the cross section for the reference flux:
\begin{equation}
    \sigma' = \frac{N(\bm\theta,\bm\phi')}{T(\bm\theta)\Phi(\bm\phi')}\text{.}
\end{equation}
It might seem, on first glance, like there is no flux error at all propagated any more,
since the only flux parameters in the equation are the \emph{constant} reference flux parameters $\bm\phi'$.
The uncertainty does however enter indirectly via the correlations of the other parameters $\bm\theta$ with the flux.

This also holds true when considering the efficiency correction as a separate step.
\autoref{eq:efficiency} can be expanded to
\begin{equation}
    N(\bm\theta, \bm\phi) = \frac{N_\text{reco}(\bm\theta,\bm\phi)}{\epsilon(\bm\theta,\bm\phi)}
    = \sum_i \frac{N_\text{reco,i}(\bm\theta) \phi_i}{\epsilon_i(\bm\theta)} \text{,}
\end{equation}
where $N_{reco,i}$ is the expected reconstructed number of events coming from the true energy bin $i$.
In this case, \autoref{eq:full-extrapolation} becomes
\begin{equation}
    N'(\bm\theta, \bm\phi) = \sum_i \frac{N_\text{reco,i}(\bm\theta) \phi_i}{\epsilon_i(\bm\theta)} \frac{\phi'_i}{\phi_i} = \frac{N_\text{reco}(\bm\theta,\bm\phi')}{\epsilon(\bm\theta,\bm\phi')} \text{,}
\end{equation}
and the cross section
\begin{equation}
    \sigma' = \frac{N_\text{reco}(\bm\theta,\bm\phi')}{\epsilon(\bm\theta,\bm\phi')T(\bm\theta)\Phi(\bm\phi')}\text{.}
\end{equation}
It seems strange that we would use the efficiency of the reference flux to correct our real data,
but in actuality we are efficiency correcting the data we \emph{would have seen} in the reference flux $N_\text{reco}(\bm\theta,\bm\phi')$, not our real data.

Lastly, we need to decide which flux to use as the reference flux.
Since this can be done \emph{after} the parameter fit,
the most logical choice is probably to use the best fit point of the flux parameters:
\begin{equation}
    \bm\phi' = \bm{\hat\phi}\text{.}
\end{equation}
The reason for this is not just that this is the best estimate for the real flux,
but it is also the point in the parameter space where any approximations and linearisations done by the fitter, like e.g. the error treatment as a covariance matrix, are most valid.
Alternatively it should also be valid to use the experiment's nominal, or \enquote{design} flux.
This would have the benefit of easier comparisons of multiple measurements done in the same neutrino beam.
On the other hand, that should probably be done with a first-approach measurement anyway.

\subsection{Multiverse unfolding}

This section will deal with the \enquote{classical} unfolding approach.
The general procedure is as follows:
\begin{itemize}
    \item Interaction models and detector simulations are used to create a detector response matrix (also called \enquote{smearing matrix}) that describes how likely an event in a certain true kinematic bin is to be reconstructed in another reconstructed kinematic bin.
    \item An unfolding algorithm is used with the recorded data and the matrix to reconstruct an estimator for the true number of events in each true kinematics bin.
    \item The smearing matrix is varied according to the systematic uncertainties of the detector and interaction models.
    Each varied matrix leads to a different unfolding and thus a different cross-section result.
    The different results can be used to build a covariance matrix.
\end{itemize}
A more in-depth example of this is described e.g. in \cite{Ruterbories:2018gub, Patrick:2018gvi}.

Let us again split the parameters into flux parameters and non-flux parameters.
The equation to calculate the cross section looks the same as in \autoref{eq:xsec-parameters}:
\begin{equation}
    \sigma = \frac{N|_{\bm\theta,\bm\phi}}{T(\bm\theta)\Phi(\bm\phi)} \text{.}
\end{equation}
In this case however, $N|_{\bm\theta,\bm\phi}$ is the unfolded (and efficiency corrected) result,
which implicitly depends on the recorded data.\footnote{We again ignore any indices or bin widths associated with a differential measurement.}
This corresponds to a first-approach measurement.
Under ideal circumstances, the flux parameters have no effect on the unfolding procedure,
and the flux uncertainty enters mainly via the integrated flux in the denominator.
If the flux shape does influence the unfolding algorithm in some way,
the correlations of flux and cross sections have to be taken into account in the model comparisons.
For this -- just like in the fitter case -- it should be easy to create a covariance matrix that correlates the flux parameters with the cross-section values by treating the flux parameters as part of the result rather than just an input.

To turn this into a second-approach measurement,
one again needs to find an extrapolation function  $N'(\bm\theta,\bm\phi)$.
Unlike in the fitter case, this presents a problem though:
The unfolding algorithm will only provide the total number of events $N$, but not the relative contribution from the different neutrino energy bins $N_i$.
To get this information from the unfolding process,
one would need to explicitly do the unfolding of the neutrino energy as well.
This is often deliberately avoided because of insufficient statistics, limited detector capabilities, and the general impossibility of measuring the neutrino energy without assuming some sort of interaction model.

Let us assume that it is possible to modify the measurement in a way to include this information though.
In this case, the extrapolation function will look the same as in \autoref{eq:extrapolation}:
\begin{equation}
    N'(\bm\theta, \bm\phi) = \sum_i N_i|_{\bm\theta,\bm\phi} \frac{\phi'_i}{\phi_i} \text{.}
\end{equation}
This time it is not possible to further simplify this though,
since the flux weights are not a direct multiplicative factor in the unfolding function.\footnote{As mentioned before, ideally the effect of the flux parameters on the unfolding should be negligible. A constant flux weight of $200\%$ over all neutrino energy bins does not change the detector response matrix. Any effect of the flux shape should also be suppressed by the analysis design.}
The second-approach cross section then becomes:
\begin{equation}
    \sigma' = \frac{N'(\bm\theta,\bm\phi)}{T(\bm\theta)\Phi(\bm\phi')}
    = \frac{\sum_i N_i|_{\bm\theta,\bm\phi} \frac{\phi'_i}{\phi_i}}{T(\bm\theta)\Phi(\bm\phi')} \text{.}
\end{equation}

One might wonder, why not just report a differential cross section over the neutrino energy in this case?
A flux-averaged measurement might still be preferable under certain circumstances,
since any model and detector uncertainties regarding the neutrino energy reconstruction only enter the result via the flux error propagation.

If the unfolded result is not available in neutrino energy bins even as an intermediary step,
the only way to extrapolate the result to a different flux is to assume a certain energy spectrum in the result.
The obvious choice is to use the Monte Carlo data that is used in the unfolding process:
\begin{align}
    N'(\bm\theta, \bm\phi)
    &= \frac{N|_{\bm\theta,\bm\phi}}{N_\text{MC}(\bm\theta,\bm\phi)} \sum_i N_{\text{MC},i}(\bm\theta, \bm\phi) \frac{\phi'_i}{\phi_i} \\
    &= \frac{N|_{\bm\theta,\bm\phi}}{N_\text{MC}(\bm\theta,\bm\phi)} \sum_i N_{\text{MC},i}(\bm\theta)\phi_i \frac{\phi'_i}{\phi_i} \\
    &= \frac{N|_{\bm\theta,\bm\phi}}{N_\text{MC}(\bm\theta,\bm\phi)} \sum_i N_{\text{MC},i}(\bm\theta)\phi'_i \\
    &= \frac{N|_{\bm\theta,\bm\phi}}{N_\text{MC}(\bm\theta,\bm\phi)} \sum_i N_{\text{MC},i}(\bm\theta, \bm\phi') \\
    &= N|_{\bm\theta,\bm\phi} \frac{N_\text{MC}(\bm\theta,\bm\phi')}{N_\text{MC}(\bm\theta,\bm\phi)} \text{.}
\end{align}
Here $N_{\text{MC}(,i)}(\bm\theta,\bm\phi)$ is the number of true events (in the $i$-th neutrino energy bin) in the simulated data assuming the given set of nuisance parameters.

So the second-approach cross section becomes:
\begin{equation}
    \sigma' = \frac{N|_{\bm\theta,\bm\phi}}{T(\bm\theta)\Phi(\bm\phi')} \frac{N_\text{MC}(\bm\theta,\bm\phi')}{N_\text{MC}(\bm\theta,\bm\phi)}\text{.}
\end{equation}
In both cases there is no \enquote{best fit} set of flux parameters available,
so the logical choice for the reference flux is the central value of the parameter throws, i.e. the nominal flux:
\begin{equation}
    \bm\phi' = \left< \bm{\phi} \right> \text{.}
\end{equation}

\subsection{Hybrid measurement}

Hybrid approaches somewhere in between the \enquote{classical} multiverse unfolding and template fitting have also been used, e.g. in \cite{Abe:2016tmq}.
Here the unfolding is repeated multiple times under differing flux and other systematic assumptions like in the multiverse approach,
but the unfolding in each case is done by doing a template fit.
The fit might, or might not include the freedom to vary the thrown parameters within varied constraints.

If the fits are \emph{not} free to vary the flux parameters,
this setup is functionally identical to the classical unfolding case,
as far as the flux uncertainty propagation is concerned.
The extrapolation can be simplified though
if the fits all use the same parametrisation for the predicted number of true events (just with different set points):
\begin{equation}
    N|_{\bm\theta,\bm\phi} = N(\bm{\hat\theta}|_{\bm\theta,\bm\phi}, \bm\phi) = \sum_i N_i(\bm{\hat\theta}|_{\bm\theta,\bm\phi}) \phi_i \text{,}
\end{equation}
\begin{equation}
    N'(\bm\theta,\bm\phi) = \sum_i N_i(\bm{\hat\theta}|_{\bm\theta,\bm\phi}) \phi_i \frac{\phi'_i}{\phi_i} = N(\bm{\hat\theta}|_{\bm\theta,\bm\phi}, \bm\phi') \text{.}
\end{equation}
Here $\bm{\hat\theta}|_{\bm\theta,\bm\phi}$ is the best-fit result of the fit under the assumptions of the thrown parameters $\bm\theta$ and $\bm\phi$.
Note that the extrapolation function is \emph{not} equal to the unfolding result at the reference flux,
as the flux assumptions of the fit are still the ones for the thrown parameters.

This method of error propagation is \emph{not} suitable
if the flux parameters are free to vary in the fit.
Depending on the relative constraint from the prior assumption and the fitted data,
the best-fit estimate of the parameters will vary much less than the data constraint allows.\footnote{If the prior constraint is much weaker than the data constraint, changing the prior's central value will not affect the $\chi^2$ surface of the parameter. The fit will always return virtually the same result.}
Thus, using only the variation of the best fit point as a measure for the uncertainty will lead to undercoverage.
Instead, the post-fit covariance matrix of the fitted parameters needs to be taken into account like in the regular template fitter case.
Since the covariance matrix will be different for every fit under different thrown systematic assumptions,
a new set of post-fit parameters will have to be drawn for each.
The cross-section uncertainty can then be deduced from the spread of all of these results.

\section{Conclusions}

We have shown the difference between reporting a flux-averaged cross-section measurement in the real flux -- a first-approach measurement -- and reporting a flux-averaged cross-section measurement in a reference flux -- a second-approach measurement.
The difference between the two is subtle,
and even if the central values of the two are identical
\drem{(e.g. if the reference flux is taken from the central value of the flux uncertainty)},
the resulting covariances can be very different.\footnote{depending on the size of the flux shape uncertainties compared to all other uncertainties in the measurement}
When used carelessly, this can lead to drawing the wrong conclusions from model comparisons to the data.

It has been shown qualitatively that in the case of the two exemplary CCQE-like measurements of T2K and MINERvA and the evaluated Genie model,
the flux shape uncertainty \drem{is} \dadd{seems to be} a subdominant but \dadd{non-negligible} \drem{important} contribution \dadd{to the total uncertainty}, which is currently not \dadd{fully} taken into account.
It is expected that in future cross-section measurements the statistical uncertainties will decrease as more data becomes available and so the relevance of the flux shape uncertainty will grow.
For these future measurements it is \drem{likely} \dadd{possible} that neglecting the flux shape uncertainty could lead to incorrect physics conclusions.
\dadd{A rigorous quantitative determination of the size of the effect would require dedicated studies by the experimental collaborations.}

First-approach measurements are somewhat simpler to implement and it is possible to perform them without assuming anything about the cross-section model, at least in principle.
They are, however, more difficult to compare with model predictions.
If the flux shape uncertainty has no influence on the result,
it is possible to vary the model predictions within those shape uncertainties and treat the resulting model uncertainty as an additional covariance on the cross-section result.
If the flux shape uncertainty on the other hand \emph{does} have an impact on the result,
the model uncertainties would need to be correlated with the reported data uncertainties.

Second-approach measurements are much simpler to compare to models.
Since they report the cross section and its uncertainty for a single, well-defined flux,
the models will only need to generate a prediction at that one flux.
The down-side to this is that it is necessary to make assumptions about the neutrino energy dependence of the cross-section to extrapolate from possible real fluxes to the reference flux.
\drem{This is only a second-order effect though, when choosing the reference flux carefully.}
\dadd{The impact of this can be minimised by choosing the reference flux carefully, e.g. using the best-fit result as reference.}
In that case, it should only introduce additional model dependence in the uncertainty propagation, not in the central value of the result.
To be conservative, the model uncertainties will have to cover many possible neutrino energy dependencies.
This means the result will lose some discrimination power when comparing it to a model with a single explicit energy dependence.

In summary:
A first-approach measurement with a correlated flux uncertainty propagation in the model would yield the better discrimination power between the two approaches,
but it requires extra effort at the time of model comparison.
The result alone is not the whole story.
A second-approach measurement is easy to compare to models.
The covariance of the result is all there is to it.
Unfortunately one loses some discrimination power due to the need of covering many potentially different energy dependencies in a single result.
When treating a first-approach result like a second-approach result by only comparing a model at a single flux,
flux shape errors are not correctly taken into account and wrong physics conclusions could be drawn.
The effect in the evaluated example analyses of T2K and MINERvA \drem{is} \dadd{seems to be} not dominant, but not negligible either.

\section*{Acknowledgements}

We would like to thank Marco Del Tutto for the very useful discussions about forward-folding analysis techniques, which in a twist of fate first lead to the (re-)discovery of the issue presented in this paper. This work was first presented and discussed within the T2K cross-section working group. We therefore are very grateful all members of the group, but are especially thankful for the useful discussions with Callum Wilkinson and Luke Pickering (your initial scepticism drove us further), Kendall Mahn and Sara Bolognesi. In the process of writing this paper we even realised that Sara had first raised this issue to T2K more than five years ago! Unfortunately she was ahead of her time and the issue did not receive the attention it deserves. Further thanks goes to Andrew Cudd, who helped ensure that we understood how the fitter works. We also state our appreciation of the T2K neutrino beamline working group for providing us with a flux covariance matrix that we could use in this paper. We further thank the MINERvA collaboration for having made the tools needed to run the studies in Sec.~\ref{sec:sizeOfEffect} publicly available and especially thank Dan Ruterbories for fruitful discussions.

\bibliography{mybibfile}

\end{document}